\documentclass[aps,preprint,prd,showpacs,nofootinbib]{revtex4}

\usepackage{latexsym}

\usepackage{latexsym}
\usepackage{amsmath}
\usepackage{graphicx}
\usepackage{subfigure}
\usepackage{dcolumn}
\usepackage{bm}
\usepackage{amssymb}
\usepackage{latexsym}
\usepackage{ulem}
\usepackage{color}
\usepackage[colorlinks,linkcolor=magenta,anchorcolor=cyan,citecolor=blue]{hyperref}

\def\be{\begin{equation}}
    \def\ee{\end{equation}}
\def\ba{\begin{eqnarray}}
    \def\ea{\end{eqnarray}}

\def\nn{\nonumber}
\def\lf{\left}
\def\rt{\right}
\def\be{\begin{equation}}
\def\ee{\end{equation}}
\def\ba{\begin{eqnarray}}
\def\ea{\end{eqnarray}}

\begin{document}

\title{Mixing of gravitational wave echoes }

\author{Zhi-Peng Li$^{1}$\footnote{lizhipeng172@mails.ucas.ac.cn}}
\author{Yun-Song Piao$^{1,2}$\footnote{yspiao@ucas.ac.cn}}

\affiliation{$^1$ School of Physics, University of Chinese Academy
of Sciences, Beijing 100049, China}



\affiliation{$^2$ Institute of Theoretical Physics, Chinese
Academy of Sciences, P.O. Box 2735, Beijing 100190, China}

\begin{abstract}

Gravitational wave (GW) echoes, if they exist, would be a probe to
the near-horizon quantum structure of black hole (BH), which has
motivated the searching for the echo signals in GW data. We point
out that the echo phenomenology related with the potential
structure might be not so simple as expected. In particular, if
the near-horizon regime of BH is modelled as a multiple-barriers
filter, the late-time GW ringdown waveform will exhibit the mixing
of echoes, even the superpositions. As a result, the amplitudes of
successive echoes might not drop sequentially.

\end{abstract}

\maketitle

\section{Introduction}

The direct detections of the gravitational wave (GW) signals by
the LIGO Scientific and Virgo Collaborations have opened up a new
window into the strong gravity regime
\cite{Abbott:2016blz,TheLIGOScientific:2017qsa}. The GW signal of
binary black holes (BHs) coalescences consists of the inspiral
phase, the merger phase and the ringdown phase. It is usually
thought that the GW ringdown signal is a powerful hint of the
existence of BH horizon. However, Cardoso et al. have pointed out
that the ringdown waveform detected is related only with the light
ring of the post-merger object rather than the horizon
\cite{Cardoso:2016rao}, see \cite{Cardoso:2017cqb} for a review.
Thus although the GW events observed are compatible with the BHs
predicted in General Relativity (GR)
\cite{TheLIGOScientific:2016src,Abbott:2018lct}, it is still
possible that the new physics might present near the horizon
\cite{Giddings:2016tla,Giddings:2017jts}.

It is well-known that the BH based on GR suffered from the
information paradox, which has inspired the modification to GR BH,
e.g.``firewalls''\cite{Almheiri:2012rt,Maldacena:2013xja},``fuzzy
ball"\cite{Mathur:2005zp}. Usually, in such modifications, as well
as in the alternatives to GR BH, e.g.
gravastar\cite{Mazur:2004fk,Visser:2003ge} (see also its origin in
early universe \cite{Wang:2018cum}), boson
star\cite{Liebling:2012fv,Brito:2017wnc,Palenzuela:2017kcg}, the
Schwarzschild horizon may be replaced by a reflective surface or
barrier.
It has been showed that if such a surface reflects GW, the
ringdown waveform of post-merger object is initially almost
similar to the ringdown signal of BH, but at late-time will show
itself a series of ``echoes"
\cite{Cardoso:2016rao,Cardoso:2016oxy,Barausse:2014tra,Abedi:2016hgu}.
see also
\cite{Price:2017cjr,Nakano:2017fvh,Mark:2017dnq,Zhang:2017jze,Bueno:2017hyj,Wang:2018mlp,Correia:2018apm,Wang:2018gin,Testa:2018bzd,Konoplya:2018yrp}
for relevant explorations.



It has been widely thought that if they exist, the echoes would be
the probes of quantum gravity physics at the near-horizon regime,
which has motivated the searching for the echo signals in GW data
\cite{Abedi:2016hgu,Maselli:2017tfq,Westerweck:2017hus,Conklin:2017lwb},
see also recent progress
\cite{Tsang:2018uie,Nielsen:2018lkf,Lo:2018sep,Wang:2019szm}.
It has been showed in Refs.\cite{Wang:2018mlp} that if the
post-merger compact object is unstable, which is collapsing into a
BH, in the post-merger ringdown waveform the echo intervals will
inevitably increase with the time, see also \cite{Chen:2019hfg}.
Relevant studies enriched the echo phenomenology and helped to the
searching for the echo signals \cite{Wang:2019szm}.



Actually, the echo waveforms related with the potential physics
near the horizon might be far complicated than expected. In a
pioneer work, Bekenstein and Mukhanov
\cite{Bekenstein:1974jk,Bekenstein:1995ju} have pointed out that
the area of BH horizon might be quantized $A\sim (integer)l_P^2$
($l_P^2$ is the Planck area). As a result, the wavelength of GWs
absorbed or emitted by a BH is also quantized, see also
\cite{Foit:2016uxn}.
Recently, in Ref.\cite{Cardoso:2019apo}, Cardoso, Foit and Kleban
have modelled the near-horizon regime of such BHs as a filter
consisting of a couple of reflective barriers, which just absorbs
the GW with some frequencies and reflects the rest, and found that
in the post-merger ringdown waveform the echo signal will be
distorted.

Inspired by Ref.\cite{Cardoso:2019apo}, we will investigate the
echo phenomenology of the BH with a multiple-barriers filter at
its near-horizon regime in details. In Sec.\ref{Numerically}, we
model our BH setup, and numerically show the ringdown waveforms of
such post-merger BH. Besides the echo signal is distorted, we
observe that the mixing and superpositions of echoes also present.
As a result, the amplitudes of successive echoes might not drop
sequentially. In Sec.\ref{Analytically}, with the Dyson series
method proposed by Correia and Cardoso \cite{Correia:2018apm}, we
analyse the corresponding ringdown waveforms. And we conclude in
Sec.\ref{Discussion}

\section{The setup}\label{Numerically}
\subsection{Near-horizon (multiple) barriers}\label{Numerically A}

\begin{figure}[htbp]
\includegraphics[scale=0.6,width=0.5\textwidth]{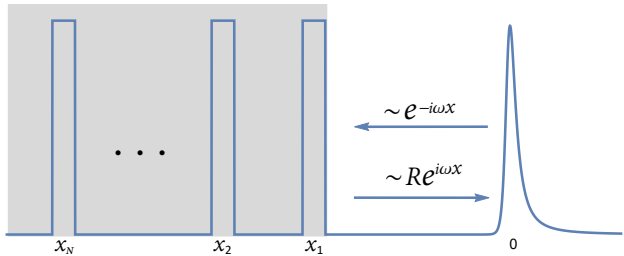}
\caption{The near-horizon structure of BH in the setup of
Sec.\ref{Numerically}. In the tortoise coordinate, the right-most
barrier is $V^{BH}(x)$ in (\ref{V}), while the effective potential
$V_{eff}(x)$ in (\ref{V}) corresponds to a set of barriers in the
shadow region. }\label{Figure1}
\end{figure}

Inspired by Ref.\cite{Cardoso:2019apo}, we model the (nonspining
for simplicity) BH as such an object, which obeys the
Schwarzschild metric
\begin{align}
ds^{2}=-B dt^{2}+{dr^{2}\over B}+r^{2}d\Omega^{2},\quad
\lf(B=1-{2M\over r}\rt)
\end{align}
at its radius $r>r_1$, but at its near-horizon regime
$(2M<r<r_1)$, where ${r_1\over 2M}-1\ll 1$, the quantum effect of
BH would bring the distinct structure, see Fig.\ref{Figure1}
plotted in the tortoise coordinate $x=r+2M\log({r\over 2M}-1)$.

In the tortoise coordinate, the Regge-Wheeler equation for the
axial gravitational perturbation is
\begin{align}
\lf[-\frac{\partial^{2}}{\partial
t^{2}}+\frac{\partial^{2}}{\partial x^{2}}-V(x)\rt]\Psi(t,x)=0 ,
\label{Phi-equation}\end{align} with \ba V(x)=
V_l^{BH}(r(x))+V_{eff}(r(x)), \label{V}\ea where $V_l^{BH}(r(x))$
is the barrier $V_l^{BH}(r)=B
\lf[\frac{l(l+1)}{r^{2}}+\frac{(1-s^2)B'}{r}\rt]$ but written in
the tortoise coordinate $x$, $l\geqslant s$ is the multipolar
index and $s=2$ \cite{Berti:2009kk}. We refer $V_l^{BH}(x)$ as the
potential barrier at the photon sphere, which is equivalent to
that of GR BH. We have put the unknown physics at the near-horizon
regime of BH into $V_{eff}(x)$, which is either a single
reflective surface, or a complicated barrier, or a set of multiple
barriers (equivalently, a special boundary condition).

The ringdown burst incident towards the horizon will be reflected
repeatedly between the barrier $V_{eff}(x)$ and $V_l^{BH}(x)$.
Thus as has been pointed out in Refs.\cite{Cardoso:2016rao,
Cardoso:2016oxy} that the ringdown waveform of post-merger BH will
consist of the primary signal, almost similar to that of GR BH,
and a series of echoes. Generally, $V_{eff}(x)$ related with the
potential physics might be not a simple reflective surface, but a
set of multiple barriers, as argued in Ref.\cite{Cardoso:2019apo}.
Relevant echo phenomenology has not yet been explored completely.

\subsection{Waveforms of echoes}

As an illustration, we will focus on the case with $l= 2$. Using
the Laplace transform $
\tilde{\Psi}(\omega,x)=\int_{0}^{\infty}\Psi(t,x)e^{i\omega t}dt$
\cite{Barausse:2014tra}, one rewrite Eq.(\ref{Phi-equation}) as
\begin{equation}
\left[\frac{\partial^{2}}{\partial
x^{2}}+\omega^{2}-V(x)\right]\tilde{\Psi}(\omega,x) ={\cal I},
\label{eq2}
\end{equation}
with ${\cal I}(\omega,x)=i\omega\Psi(0,x)-{\dot\Psi}(0,x)$, where
$\Psi(0,x)$ and ${\dot\Psi}(0,x)$ are the initial conditions of
$\Psi(t,x)$. We consider $V_{eff}(x)$ in (\ref{V}) as \be
V_{eff}(x)=2\sum_{j=1}^N V_j \delta(x-x_j),\label{Vdelta}\ee where
$2V_j$ is the height of Delta barrier at $x_j$, which is a
simplified model of multiple barriers, but is sufficient to catch
the echo phenomenology of the setup depicted in Fig.\ref{Figure1}. Solving out
$\tilde{\Psi}(\omega,x)$, we may get $\Psi(t,x)$ by
\begin{equation}
\Psi(t,x)={1\over
2\pi}\int_{-\infty}^{\infty}\tilde{\Psi}(\omega,x)e^{-i\omega
t}d\omega.
\end{equation}

\begin{figure}[htbp]
    \includegraphics[scale=2,width=0.48\textwidth]{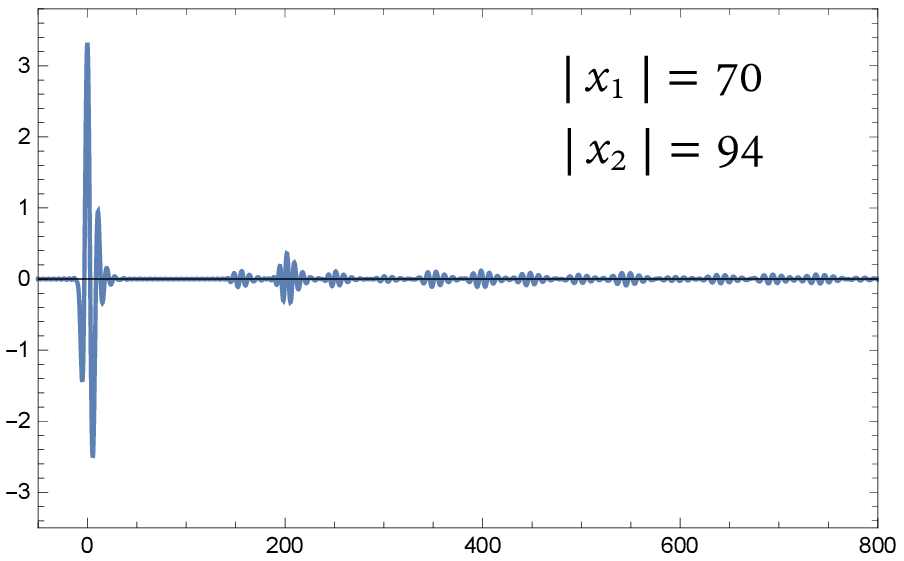}
    \includegraphics[scale=2,width=0.48\textwidth]{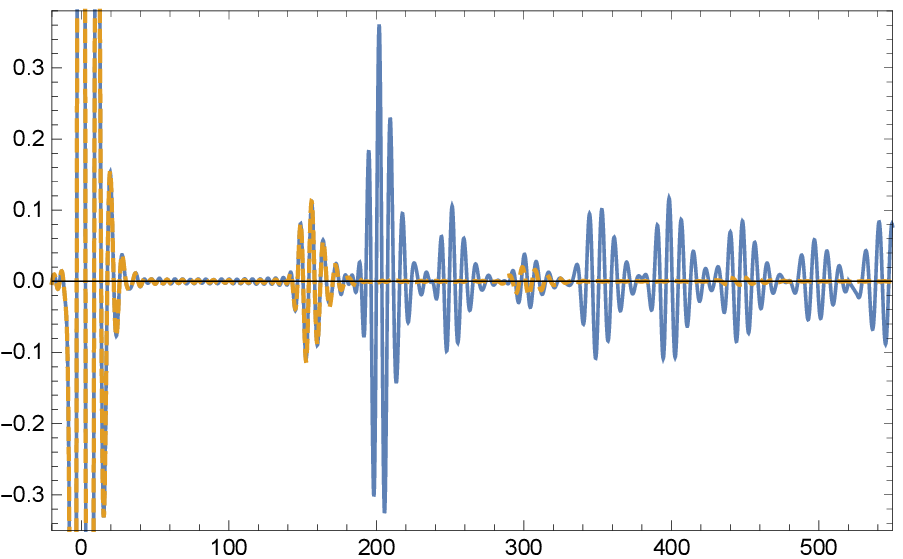}
    \caption{Ringdown waveforms with respect to $t$. The right panel is the
    magnification of the left panel. As a comparison, the yellow curve in the right panel is
    the ringdown waveform with a single Delta barrier near the horizon ($|x_1|=70$).}
    \label{echo_L70_l24.eps}
\end{figure}

\begin{figure}[htbp]
    \includegraphics[scale=2,width=0.47\textwidth]{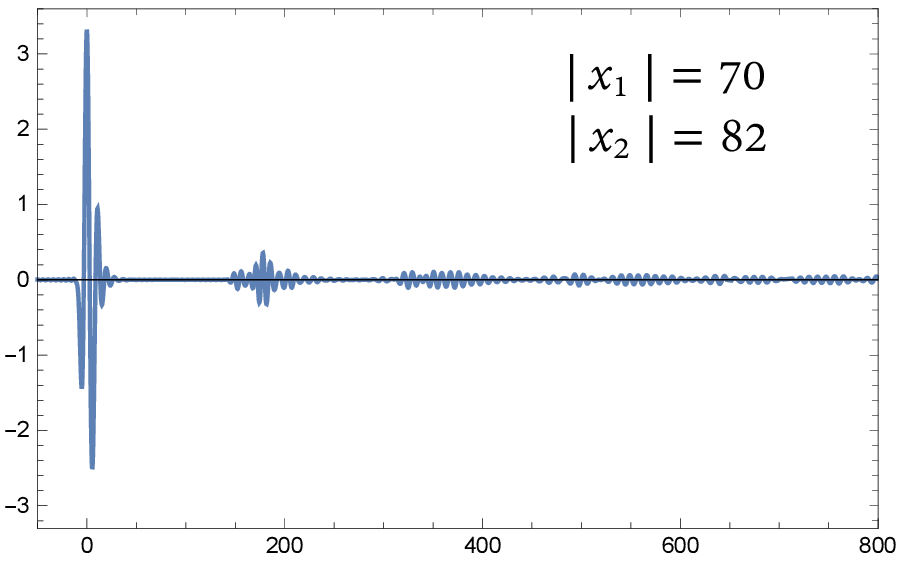}
    \includegraphics[scale=2,width=0.48\textwidth]{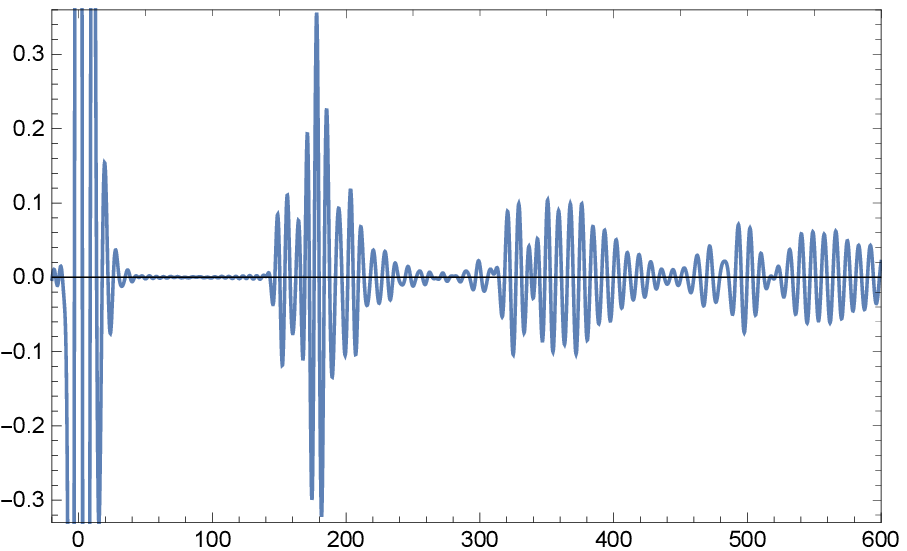}
    \caption{Ringdown waveforms with respect to $t$. The right panel is the
    magnification of the left panel.}
    \label{echo_L70_l12.eps}
\end{figure}

\begin{figure}[htbp]
    \includegraphics[scale=2,width=0.48\textwidth]{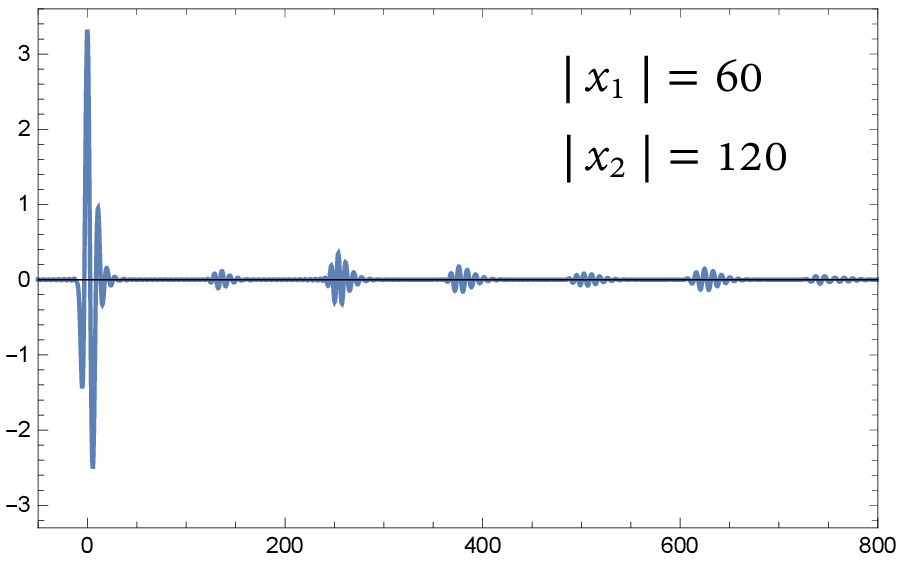}
    \includegraphics[scale=2,width=0.48\textwidth]{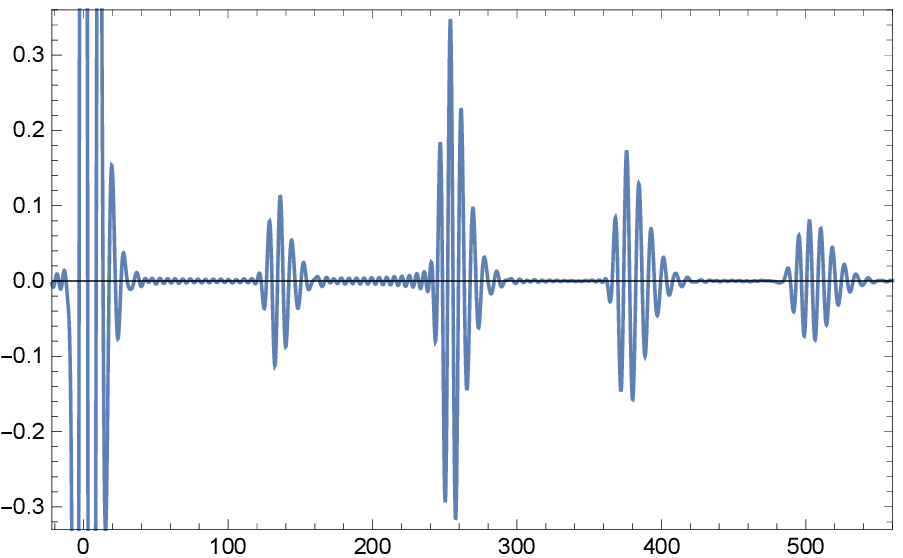}
    \caption{Ringdown waveforms with respect to $t$ in the model with $|x_2-x_1|=|x_1|$. The right panel is the
        magnification of the left panel. }
    \label{echo_L60_l60.eps}
\end{figure}

\begin{figure}[htbp]
    \includegraphics[scale=2,width=0.48\textwidth]{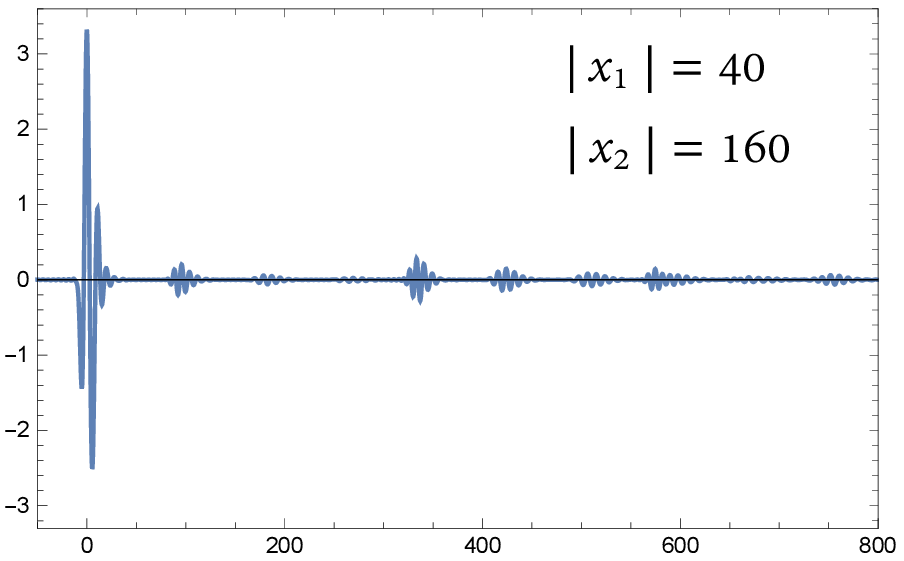}
    \includegraphics[scale=2,width=0.48\textwidth]{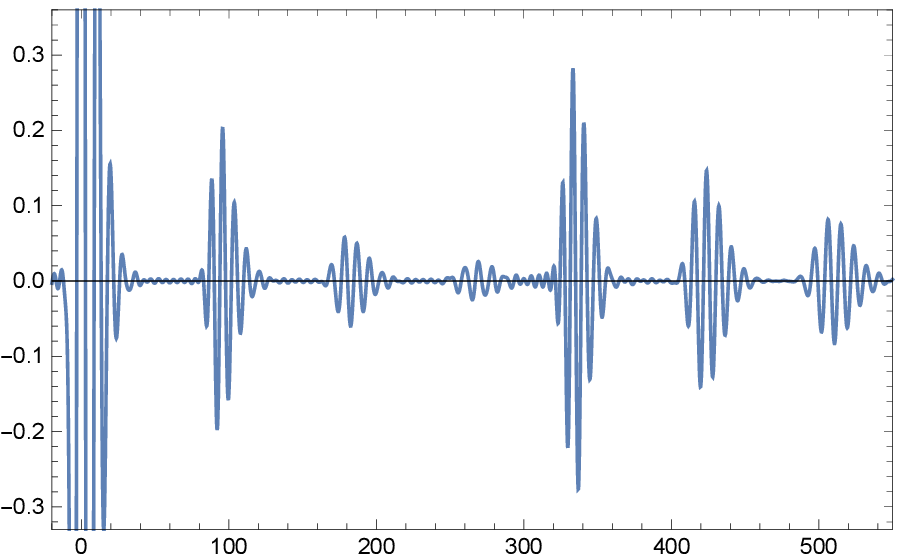}
    \caption{Ringdown waveforms with respect to $t$ in the model with $|x_2-x_1|=3|x_1|$. The right panel is the
        magnification of the left panel.}
    \label{echo_L40_l120.eps}
\end{figure}

We set
\begin{align}
\Psi(0,x)=e^{-(x-5)^{2}/9},\quad\, {\partial_t\Psi}(0,x)=0,
\label{initial}
\end{align}
($2M=1$ for convenience) and also $N=2$ in (\ref{Vdelta}), i.e.
$V_{eff}(x)=2\sum_{j=1}^2 V_j \delta(x-x_j)$ consists of a couple
of Delta barriers, and plot the corresponding ringdown waveforms
$\Psi(t,x)$ in Figs.\ref{echo_L70_l24.eps},
\ref{echo_L70_l12.eps}, \ref{echo_L60_l60.eps} and
\ref{echo_L40_l120.eps} for different values of $x_1$ and $x_2$,
respectively, where the height of barriers are $2V_1=0.5$ or $1$,
$2V_2=60$.

We see in Fig.\ref{echo_L70_l24.eps} that after the primary
ringdown burst a series of echoes present, as expected, but the
intervals of successive echoes seems not to equal. Here, the case
is different from that in Ref.\cite{Wang:2018mlp}, where the
change of echo intervals is caused by the shift of reflective
surface or barrier. In addition, as showed in
Ref.\cite{Cardoso:2019apo}, when $x_2$ is very close to $x_1$, so
that $2|x_2-x_1|\lesssim \Delta_{echo}$, where $\Delta_{echo}$ is
the width of echo waveform, the echoes will be significantly
distorted, see Fig.\ref{echo_L70_l12.eps}.

We see in Fig.\ref{echo_L60_l60.eps} that for $|x_2-x_1|=|x_1|$,
the echoes reflected by $V_{eff}(x)$ will present at
$t_n-t_{ringdown}=2n|x_1|$, but the echoes exhibit certain
superposition and cancellation, which occur at
$t_n-t_{ringdown}=2n|x_1|$ ($n\geqslant 2$), so that the
amplitudes of successive echoes seem be out of order. The
amplitude of echo after the superposition will be amplified.
Actually, even if $|x_2-x_1|=|x_1|$, in the physical coordinate
$r$ the barrier $V_1$ is far closer to $V_2$ than the barrier
$V^{BH}(r)$. Similarly, in Fig.\ref{echo_L40_l120.eps}, we set
$|x_2-x_1|=3|x_1|$, and see that the superposition and
cancellation occur at $t_n-t_{ringdown}=2n|x_1|$ ($n\geqslant 4$).

We also plot the ringdown waveforms for $V_{eff}(x)$ with $N=3$ in
Appendix A. The waveforms are qualitatively similar to the cases
with $N=2$.

\section{Analytic studies}\label{Analytically}

\subsection{Boundary conditions}

It is interesting to have an insight into the echo phenomenology
showed in Sect.\ref{Numerically A} by analytically solving
Eq.(\ref{eq2}). We first set the boundary conditions.

The wave $ \tilde{\Psi}$ should obey the outgoing wave condition
\be \tilde{\Psi}\sim e^{i\omega x}\label{outgoing}\ee  as $
x\rightarrow\infty $. As imagined in Sect.\ref{Numerically}, the
physics of GR BH has been modified at $r<r_1$ (equivalently
$x<x_1$ in the tortoise coordinate). However, in despite of what
the potential barrier $V_{eff}(x)$ looks like, one always may
regard the surface at $x_1$ as an effective boundary, near which
$\tilde{\Psi}$ should satisfy \cite{Mark:2017dnq}
\begin{equation}
\tilde{\Psi}\sim e^{-i\omega x}+R_{eff}(\omega)e^{i\omega x},
\label{reflect}\end{equation} where $R_{eff}(\omega)$ is the
effective reflection coefficient (RC). In particular, one has
$R_{eff}(\omega)=0$ for the GR BH (without the barrier
$V_{eff}(x)$), and $R_{eff}(\omega)=-e^{2i\omega |x_1|}$ for the
boundary condition ${\tilde \Psi}(\omega,x_1)=0$ at $x_1$. Thus in
certain sense, $R_{eff}(\omega)$ actually encodes the internal
structure of $V_{eff}(x)$.

Considering that $V_{eff}(x)=2\sum_{j=1}^2 V_j \delta(x-x_j)$
consists of a couple of barriers, we have \ba
R_{2,eff}=R_{1}+R_{2}T_{1}^{2}
\sum_{m=0}^{\infty}(\tilde{R}_{1}R_{2})^{m}
=R_{1}+\frac{R_{2}T_{1}^{2}}{1-\tilde{R}_{1}R_{2}},
\label{Reff}\ea where the integer $m$ corresponds to the roundtrip
number of GW between the barriers $V_1$ and $V_2$. Here, for the
Delta barrier with the height $2V_j (j=1,2\cdots)$, \be
R_{j}=-\frac{V_{j}}{V_{j}-i\omega}e^{-2i\omega x_{j}},\quad
\tilde{R}_{j}=-\frac{V_{j}}{V_{j}-i\omega}e^{2i\omega x_{j}},\quad
T_{j}=-\frac{i\omega}{V_{j}-i\omega}. \label{RT}\ee $R_{j}$ is the
RC of a single barrier $V_j$ for the wave incident from right,
while $\tilde{R}_{j} $ is that from left, $T_j$ is the TC of a
single barrier $V_{j}$. Thus we have \be {\tilde
R}_{1}R_{2}=\left( \frac{V_{1}}{V_{1}-i\omega}\right)\left(
\frac{V_{2}}{V_{2}-i\omega}\right)e^{2i\omega (x_{1}-x_2)}.
\label{R1R2}\ee

Considering that $V_{eff}(x)=2\sum_{j=1}^3 V_j \delta(x-x_j)$,
i.e.3-Delta barrier, we have the effective RC ${R}_{3,eff}$ as \be
{R}_{3,eff}={ R}_{2,eff}+\frac{R_{3}{T}_{2,eff}}{1-\tilde{
R}_{2,eff}R_{3}}, \label{R3eff}\ee where $R_{2,eff}$ is the
effective RC (\ref{Reff}) of 2-Delta barriers, $\tilde{R}_{2,eff}$
is that for the wave incident from left, and $R_3$ is the RC of a
single Delta barrier at $x_3$, see (\ref{RT}), $T_{2,eff}$ is the
effective TC, see Appendix B for the expressions of
$\tilde{R}_{2,eff}$ and $T_{2,eff}$. In Appendix B, we also
verified Eqs.(\ref{Reff}) and (\ref{R3eff}).


The system with 3-Delta barriers actually corresponds to that with
2-barrier (the barriers $V_3$ and $V_{2,eff}$). Thus with the
replacements $R_1\rightarrow R_{2,eff}$, $T_1\rightarrow
T_{2,eff}$ and $R_2\rightarrow R_3$ in Eq.(\ref{Reff}), we will
immediately get Eq.(\ref{R3eff}). It is not difficult to
straightly write out the effective RC of $j$-Delta barriers \be
{R}_{j,eff}={
R}_{(j-1),eff}+\frac{R_{j}{T}_{(j-1),eff}^{2}}{1-\tilde{
R}_{(j-1),eff}R_{j}}. \label{RNeff}\ee Thus with (\ref{RT}) and
the recursive relationship (\ref{RNeff}), the full result of the
effective RC ${R}_{N,eff}$ of $N$-Delta barriers may be worked
out.

Physically, Eq.(\ref{eq2}) is equivalent to
\begin{equation}
\left[\frac{\partial^{2}}{\partial
    x^{2}}+\omega^{2}-V^{BH}(x)\right]\tilde{\Psi}(\omega,x) ={\cal I},
\label{req2}
\end{equation}
with the boundary condition (\ref{outgoing}) and (\ref{reflect}),
where $V_{eff}(x)$ (related with the near-horizon physics of BH)
in (\ref{V}) has been encoded in $R_{eff}(\omega)$.



\subsection{Review on the Dyson series method}


To analyse the ringdown waveforms, we will apply the Dyson series
method proposed in Ref.\cite{Correia:2018apm}. Here, we briefly
review it.

Defining the operator ${\cal A}=\frac{\partial^{2}}{\partial
x^{2}}+\omega^{2}$, we have ${\cal A}^{-1}\tilde{\Psi}=\int
g(x,x^\prime)\tilde{\Psi}dx'$, where Green's function
$g(x,x^\prime)$ satisfying the boundary conditions
(\ref{outgoing}) and (\ref{reflect}) is \ba
g(x,x^\prime)=\frac{e^{i\omega|x-x'|}+R_{eff}e^{i\omega(x+x')}}{2i\omega}.
\label{Green}\ea Here, $R_{eff}$ is the effective RC of
$V_{eff}(x)$, and for a couple of Delta barriers, $R_{eff}$ equals
to $R_{2,eff}$ in Eq.(\ref{Reff}). According to (\ref{Green}),
$g(x,x^\prime)$ may be separated into $g=g_{O}+{R}_{eff}g_{R,eff}
$ with
\begin{equation}
g_{O}(x,x')=\frac{e^{i\omega|x-x'|}}{2i\omega},\quad g_{{
R,eff}}(x,x')=\frac{e^{i\omega(x+x')}}{2i\omega},
\end{equation}
which correspond to that of open system without $V_{eff}(x)$ and
that reflected by $V_{eff}(x)$, respectively. As a result, we have
${\cal A}^{-1}={\cal A}^{-1}_{O}+{\cal A}^{-1}_{R,eff}$ with \be
{\cal A}^{-1}_O\tilde{\Psi}=\int
g_O(x,x^\prime)\tilde{\Psi}dx',\quad {\cal
A}^{-1}_{R,eff}\tilde{\Psi}=R_{eff}\int g_{
R,eff}(x,x^\prime)\tilde{\Psi}dx', \ee

We rewrite Eq.(\ref{req2}) as $\tilde{\Psi} ={\cal A}^{-1}{\cal
I}+{\cal A}^{-1}\hat{V}^{BH}\tilde{\Psi}$, which is \ba
\tilde{\Psi} &=&\lf({\cal A}^{-1}_{O}+{\cal
A}^{-1}_{R,eff}\rt){\cal I}+\lf({\cal A}^{-1}_{O}+{\cal
A}^{-1}_{R,eff}\rt)\hat{V}^{BH}\tilde{\Psi}. \label{Dyson}\ea The
Dyson series solution of Eq.(\ref{Dyson}) is \be
\tilde{\Psi}=\tilde{\Psi}_{O}+\sum_{n=1}^{\infty}\tilde{\Psi}_{n}
\ee with  \be \tilde{\Psi}_{O}=\sum_{p=0}^{\infty}\lf({\cal
A}_O^{-1}\hat{V}^{BH}\rt)^p \lf({\cal A}_O^{-1}{\cal I}\rt), \ee
\be \tilde{\Psi}_{n}=\sum_{p=n}^{\infty}\sum_{\{p,n\}}\lf({\cal
A}_{O}^{-1}\rt)^{p-n}\lf({\cal A}^{-1}_{R,eff}\rt)^n
\lf(\hat{V}^{BH}\rt)^{p-1}{\cal I},\label{Psin}\ee where the sum $\sum_{\{p,n\}} $ only includes all distinct
possibilities of ordering $ n $ $ {\cal A}^{-1}_{R,eff} $ in
$ p $ spots.

We actually have the infinite number of Dyson series with the
barrier $V^{BH}$ at the photon sphere. Here, $\tilde{\Psi}_{O}$ is
the waveform of open system without the effective barrier
$V_{eff}$, which is irrelevant with ${\cal A}_{R,eff}^{-1}$, while
the ``reflected" waveform $\tilde{\Psi}_{n}\sim ({\cal
A}^{-1}_{R,eff})^n\sim R_{eff}^n$, where $n$ is the roundtrips
number of wave between the barriers $V_{eff}(x)$ and $V^{BH}(x)$.
The formal solution (\ref{Psin}) is actually equivalent to \ba
\tilde{\Psi}_{n}(\omega,x)&=&{R}^{n}_{eff}\sum_{p=n}^{\infty}\int
\sum_{\{p,n\}} g_{R,eff}(x,x^\prime_{1}) \cdots
g_{R,eff}(x^\prime_{n-1},x^\prime_{n})\nn\\ &&
g_{O}(x^\prime_{n},x^\prime_{n+1})\cdots
g_{O}(x^\prime_{p-1},x^\prime_{p}) V^{BH}(x^\prime_{1})\cdots
V^{BH}(x^\prime_{p-1}){\cal
I}(\omega,x^\prime_{p})dx^\prime_{1}\cdots dx^\prime_{p},
\label{Psinn}\ea which is a multiple integrals.

We will only be interested in the waveform $\tilde{\Psi}_{n}$. The
GR barrier $V^{BH}$ at the photon sphere must be speculated to
calculate $\tilde{\Psi}_{n}$. However, it is not required for our
analysis. Here, we assumed $V^{BH}=2V_0\delta(x)$ for simplicity
(but without loss of physics we care). Now, the multiple integrals
in (\ref{Psinn}) may be reduced to one integral only for
$x_p^\prime=x^\prime$. The corresponding $\tilde{\Psi}_{n}$ has
been calculated in Ref.\cite{Correia:2018apm}, \ba
\tilde{\Psi}_{n}(\omega,x)={R}^{n}_{eff}\int  dx' {\cal
I}(\omega,x') \lf[R_{0}^{n-1}{e^{i\omega(x+x')}\over {2i\omega}}
+R_{0}^{n+1}{e^{i\omega(|x|+|x'|)}\over {2i\omega}} +
R_{0}^{n}{e^{i\omega(x+|x'|)}+e^{i\omega(x'+|x|)}\over
{2i\omega}}\rt], \label{Psin3}\ea where \begin{equation}
R_{0}(\omega)=-\frac{V_{0}}{V_{0}-i\omega}
\end{equation} is the RC of a single
Delta barrier at $x_0=0$, see (\ref{RT}).

\subsection{Mixing of echoes}

We will analyse the mixing of echoes with $N=2$ in (\ref{Vdelta}),
i.e. 2-Delta barriers, as example. The cases with $N\geqslant 3$
are similar.

According to Eq.(\ref{Psin3}), the late-time ringdown waveform is
closely related with ${R}^{n}_{eff}$. With (\ref{RT}), we rewrite
$R_{eff}$ in Eq.(\ref{Reff}) as
\begin{align}
{R}_{eff}=R_{1}\lf[1+\left(\frac{i\omega}{V_{1}}\right)^{2}\sum_{m=1}^{\infty}({\tilde
R}_{1}R_{2})^{m}\rt].
\end{align}
We have \begin{align} R_{eff}^n
=& R_{1}^{n} \sum_{k=0}^{n}\left(
\begin{array}{ccc} n\\k
\end{array}\right)\left(\frac{i\omega}{V_{1}}\right)^{2k}\lf[\sum_{m=1}^{\infty}({\tilde R}_{1}R_{2})^{m}\rt]^{k}\nonumber\\
=& R_{1}^{n}\sum_{k=0}^{n}\left(
\begin{array}{ccc} n\\k
\end{array}\right)\left(\frac{i\omega}{V_{1}}\right)^{2k}\sum_{k_{m}\geq0,\sum_m k_{m}=k}\frac{k!}{k_{1}!k_{2}!\cdots k_{m}!\cdots }\prod_{m=1}^{\infty}\lf[({\tilde
R}_{1}R_{2})^{m}\rt]^{k_{m}}.
\label{Reffn3}\end{align}
We explain it as follows. When $k=0$, all $k_m$ must satisfy
$k_m=0$, so $R_{eff}^n=R_1^n$. When $k=1$, all $k_{m\neq
m^\prime}=0$ except for $k_{m^\prime}=1$ ($m^\prime$ runs from $1$
to $\infty$), we have \be R_{eff}^n\sim
R_1^n\sum_{k_{m}\geq0,\sum_m
k_{m}=k}\prod_{m=1}^{\infty}\lf[({\tilde
R}_{1}R_{2})^{m}\rt]^{k_{m}}=R_1^n\sum_{m=1}^{\infty}({\tilde
R}_{1}R_{2})^{m}.\ee  Similarly, when $k=2$, either all $k_{m\neq
m^\prime,m^{\prime\prime}}=0$ except for $k_{m^\prime}=1$,
$k_{m^{\prime\prime}}=1$ ($m^\prime\neq m^{\prime\prime}$), or all
$k_{m\neq m^\prime}=0$ except for $k_{m^\prime}=2$.

Combining (\ref{Reffn3}) and the solution (\ref{Psin3}) of
${\tilde \Psi}_n$, we have the ${\tilde n}$th echo (in the
frequency-domain) of the ringdown burst as \ba
\tilde{\Psi}_{\tilde n}(\omega,x)&\sim & R_1^n
\prod_{m=1}^{\infty}\lf[({\tilde
R}_{1}R_{2})^{m}\rt]^{k_{m}}e^{i\omega
x}\nonumber\\
&=& e^{i\omega x}\exp{\lf[-2i\omega n
x_1-2i\omega(x_2-x_1){\sum_{m=1}^{\infty} mk_{m}}\rt]}
\label{nprime}\ea for $\sum_m k_m=k\leqslant n$, where ${\tilde
R}_{1}R_{2}$ is showed in (\ref{R1R2}). We eventually get
${\tilde\Psi}_{\tilde n}(t,x)$ through $\Psi_{\tilde n}(t,x)\sim
\int\tilde{\Psi}_{\tilde n}(\omega,x)e^{-i\omega t}d\omega$.
According to Eq.(\ref{nprime}), it is found that after the primary
ringdown signal at $t_{ringdown}$, the $\tilde n$th echo will
appear at \be t_{\tilde n}-t_{ringdown}=2\lf[n
|x_1|+|x_1-x_2|{\sum_{m=1,\sum_m k_m=k\leqslant n}^{\infty}
mk_{m}}\rt].\label{tn}\ee  When $m=0$, we have $t_{\tilde
n}-t_{ringdown}=2n |x_1|$, the ordinal number ${\tilde
n}=1,2,3\cdots$ of echoes coincides with $n$, which is just the
well-known result \cite{Cardoso:2016rao} in single barrier model.
The case with the multiple barriers is different. We see that for
a fixed $n$, a series of ``echoes"  (so-called the sub-echo) of
the $n$th echo will also present with the intervals \be \Delta
t_{k,k_m}=2|x_1-x_2|{\sum_{m=1,\sum_m k_m=k\leqslant n}^{\infty}
mk_{m}}\ee after the $n$th echo. Thus the ordinal number ${\tilde
n}=1,2,3\cdots$ of echoes will be arranged by not only $n$, but
$k$, $m$, $x_1$ and $x_2$. This suggests that the echoes are
mixing.

The first (${\tilde n} =1$) echo presents at
$t_{1}-t_{ringdown}=2|x_1|$, which corresponds to $k=0$ (all
$k_m=0$) for $n=1$ in Eq.(\ref{tn}). When $|x_1-x_2|< |x_1|$, the
second (${\tilde n} =2$) echo will present at
$t_{2}-t_{ringdown}=2|x_1|+2|x_1-x_2|$, which corresponds to $k=1$
(all $k_m=0$ except for $k_1=1$) for $n=1$ and actually is the
sub-echo of the $(n=1)$th echo. Its second sub-echo is at
$t-t_{ringdown}=2|x_1|+4|x_1-x_2|$, which also corresponds to
$k=1$ but all $k_m=0$ except for $k_2=1$, see
Fig.\ref{echo_L70_l24.eps} in Sect.IIB. Moreover, if $|x_1-x_2|\ll
|x_1|$, we will have $t_{2}-t_{ringdown}\simeq t_{1}-t_{ringdown}$
(the barrier $V_2$ is very close to $V_1$, so that
$2|x_2-x_1|\lesssim \Delta_{echo}$), the first echo will be
significantly affected by its sub-echo, and distorted, as found in
Ref.\cite{Cardoso:2019apo}, see also Fig.\ref{echo_L70_l12.eps} in
Sect.IIB.

\begin{figure}[htbp]
    \includegraphics[scale=2,width=0.49\textwidth]{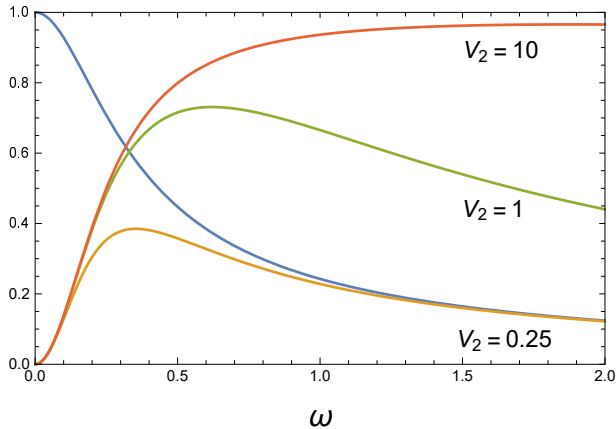}
\caption{The blue curve is the amplitude ${\cal A}_{{\tilde n}=1}$
of the first echo with $V_1=0.25$, and the others are ${\cal
A}_{{\tilde n}=2}$ with $V_1=0.25$, $V_2=0.25,1,10$, respectively.}
    \label{A2_A1}
\end{figure}

Generally, it is hardly possible that the amplitude ($\sim
|R_1|^2$) of echo with $n=2$ is larger than that ($\sim |R_1|$)
with $n=1$, since $|R_1|\leqslant 1$. Thus in the single barrier
model, the amplitudes of successive echoes drop sequentially.
However, when the multiple barriers are considered, the case is
altered. It is possible that the amplitude ${\cal A}_{{\tilde
n}=2}\sim |R_2T_1^2|$ of the second echo is larger than ${\cal
A}_{{\tilde n}=1}\sim |R_1|$. We plot ${\cal A}_{{\tilde n}=2}$
and ${\cal A}_{{\tilde n}=1}$ with respect to $\omega$ in
Fig.\ref{A2_A1}, respectively, and see that ${\cal A}_{{\tilde
n}=2}>{\cal A}_{{\tilde n}=1}$ is possible, if $V_2>V_1$.

Another possibility causing the anomalies of echo amplitudes is
that if \be n |x_1|+|x_1-x_2|{\sum_{m=1,\sum_m k_m=k\leqslant
n}^{\infty} mk_{m}}= n^\prime
|x_1|+|x_1-x_2|{\sum_{m^\prime=1,\sum_{m^\prime}
k^\prime_{m^\prime}=k^\prime\leqslant n^\prime}^{\infty} m^\prime
k^\prime_{m^\prime}}, \ee  the echoes with different $n$ will
superpose each other. The superposition of echoes is the unique
phenomenon happening only in the multiple barriers models.

In Fig.\ref{echo_L60_l60.eps} of Sect.IIB, where
$|x_1|=|x_1-x_2|$, we see that the second echo presents at
$t_{2}-t_{ringdown}=4|x_1|$, which is just the superposition of
the $(n=2)$th echo with the sub-echo ($n^\prime=1,k^\prime=1,$ all
$k^\prime_{m^\prime}=0$ except for $k^\prime_1=1)$ of the
$(n=1)$th echo. And so on, the third echo presents at
$t_{3}-t_{ringdown}=6|x_1|$, which is the superposition of the
$(n=3)$th echo with the sub-echo $(n^\prime=2,k^\prime=1$, all
$k^\prime_{m^\prime}=0$ except for $k^\prime_1=1)$ and the
sub-echo $(n^{\prime\prime}=1,k^{\prime\prime}=1,$ all
$k^{\prime\prime}_{m^{\prime\prime}}=0$ except for
$k^{\prime\prime}_2=1)$.

\section{Discussion}\label{Discussion}

Quantum gravity physics, invoked by the information paradox of BH,
might result in some microstructure at the near-horizon regime of
BH, which may reflect GWs. The GW echoes, if they exist, will be a
promising probe of such physics, which has excited the searching
for the echo signals in GW data.



However, the potential physics responsible for the GW echoes is
actually unknown, which is still in the exploration, so the echo
phenomenology might be not so simple as expected. Inspired by
Ref.\cite{Cardoso:2019apo}, we show that if the near-horizon
regime of BH is modelled as a multiple-barriers filter with
different spacings between barriers, the GW ringdown waveform of
post-merger BH will exhibit the mixing of echoes, even the
superpositions. As a result, the echo amplitudes might not drop
sequentially.


Though the post-merger BH we considered is a nonspinning BH,
extending it to the Kerr BH is straight. The effect of ergoregion
in Kerr BH on the GW echoes has been studied in
Refs.\cite{Vicente:2018mxl,Barausse:2018vdb}. In addition, it is
interesting to present a full ``template" of the ringdown
waveforms with the mixing and superpositions of echoes, along the
line in Refs.\cite{Testa:2018bzd}. It is also interesting to check
what a stochastic GW background such echoes will result in, which
might be also substantially detectable \cite{Du:2018cmp}.

It is actually possible to find the GW echoes, as more and more GW
events with higher signal-to-noise ratio are detected
\cite{Cardoso:2017cqb}. Our work suggests that the echo
phenomenology related with the potential physics might be far
richer than expected, so identifying relevant signals will be a
more challenging task.

\textbf{Acknowledgments}

We would like to thank the cooperation with Jun Zhang and
Shuang-Yong Zhou at the initial stage of this project, and also
thank Yu-Tong Wang for discussion. This work is supported by NSFC,
Nos.11575188,11690021.

\appendix
\section{The echo waveforms for $N=3$}

We plot the echo waveforms for $V_{eff}(x)$ with $N=3$ in this
Appendix.

\begin{figure}[htbp]
    \includegraphics[scale=2,width=0.47\textwidth]{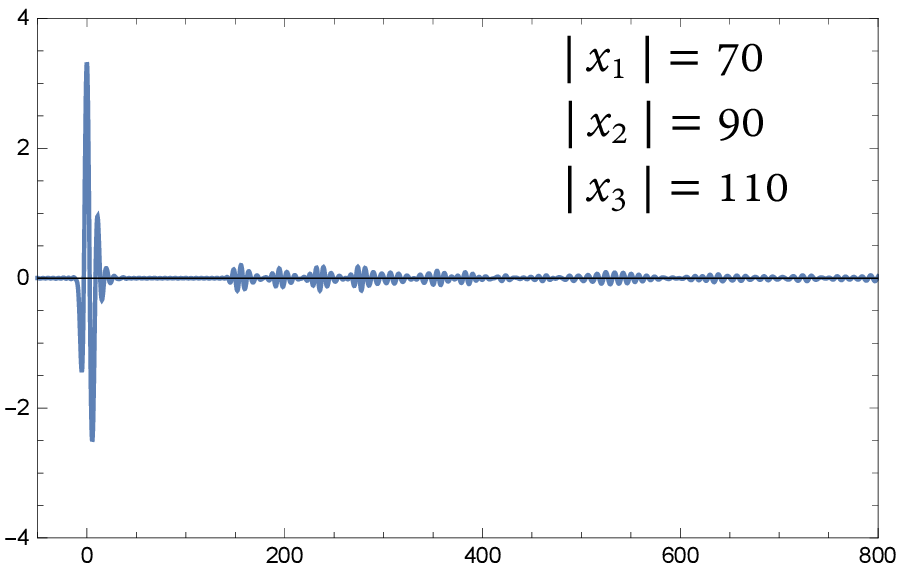}
    \includegraphics[scale=2,width=0.48\textwidth]{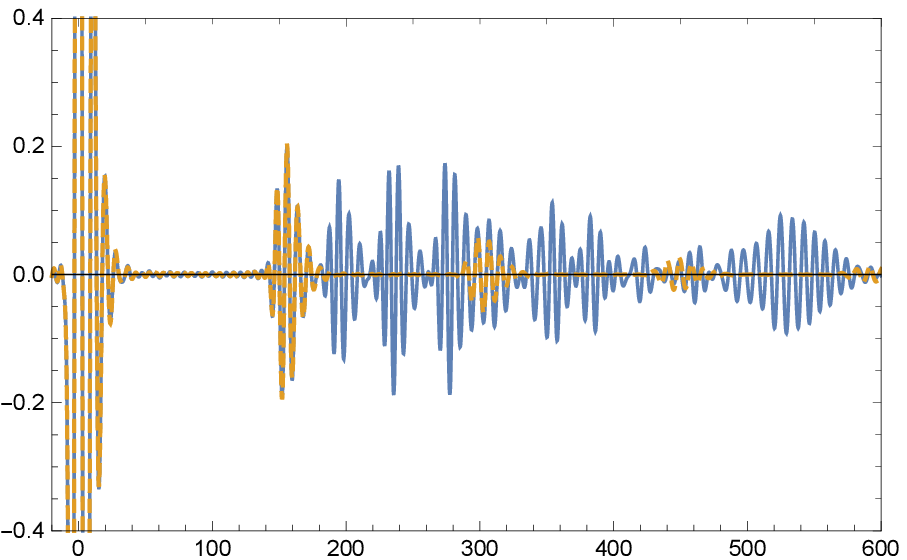}
    \caption{Ringdown waveforms with respect to $t$. The right panel is the
    magnification of the left panel. As a comparison, the yellow curve in the right panel is
    the ringdown waveform with a single Delta barrier near the horizon ($|x_1|=70$).}
    \label{}
\end{figure}

\begin{figure}[htbp]
    \includegraphics[scale=2,width=0.48\textwidth]{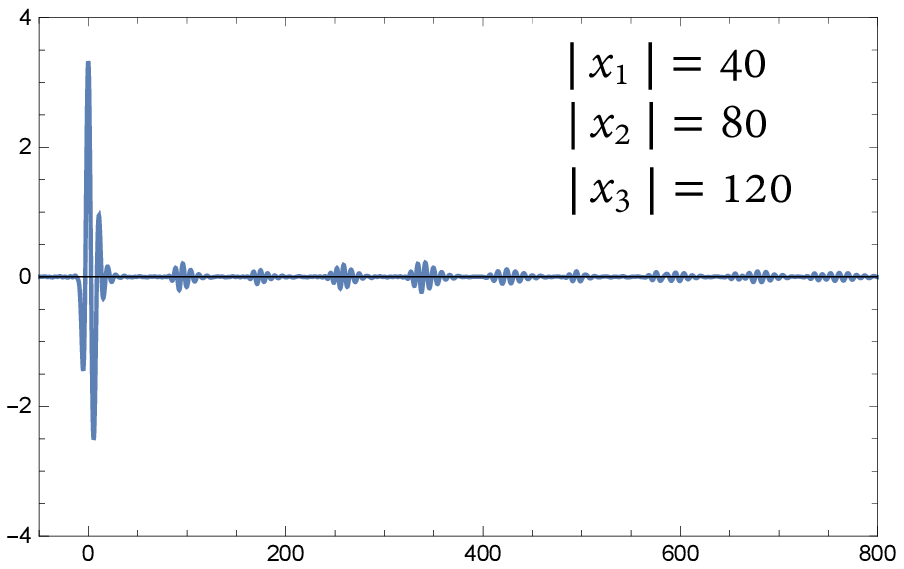}
    \includegraphics[scale=2,width=0.48\textwidth]{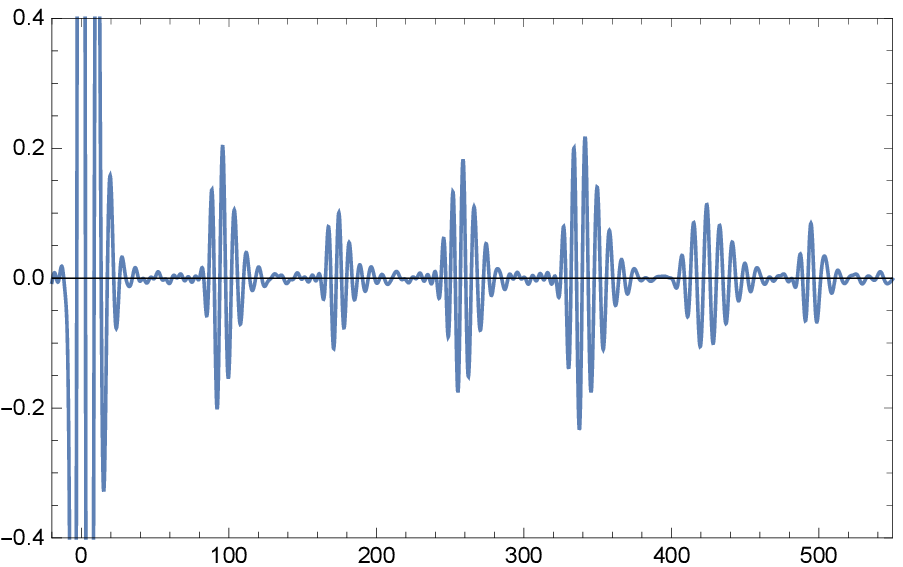}
    \caption{Ringdown waveforms with respect to $t$ in the model with $|x_3-x_1|=|x_2-x_1|=|x_1|$. The right panel is the
    magnification of the left panel.}
    \label{}
\end{figure}

\section{The effective RC for $N=2,3$}

In this Appendix, we will briefly verify Eqs.(\ref{Reff}) and
(\ref{R3eff}).

We consider $N=2$ first. The Delta barriers at $x_1$ and $x_2$
divide the $x$-space into the regions I, II, III. The waves
${\tilde \Psi}_I=e^{-i\omega x}+A_{1}e^{i\omega x}$, ${\tilde
\Psi}_{II}=A_{2}e^{-i\omega x}+A_{3}e^{i\omega x}$ and ${\tilde
\Psi}_{III}=A_{4}e^{-i\omega x}$ in corresponding regions must
satisfy
\begin{align}
&e^{-i\omega x_{1}}+A_{1}e^{i\omega x_{1}}=A_{2}e^{-i\omega
x_{1}}+A_{3}e^{i\omega x_{1}}, \\ &A_{2}e^{-i\omega
x_{2}}+A_{3}e^{i\omega x_{2}}=A_{4}e^{-i\omega x_{2}}, \\
&-i\omega e^{-i\omega x_{1}}+i\omega A_{1}e^{i\omega
x_{1}}+i\omega A_{2}e^{-i\omega x_{1}}-i\omega A_{3}e^{i\omega
x_{1}}=2V_{1}\left( e^{-i\omega x_{1}}+A_{1}e^{i\omega
x_{1}}\right), \\ &-i\omega A_{2}e^{-i\omega x_{2}}+i\omega
A_{3}e^{i\omega x_{2}}+i\omega A_{4} e^{-i\omega
x_{2}}=2V_{2}A_{4}e^{-i\omega x_{2}},
\end{align}
at $x_1$ and $x_2$. The effective RC is \be
R_{2,eff}=A_{1}=\frac{V_{2}(V_{1}+i\omega)e^{-2i\omega
x_{2}}-V_{1}(V_{2}-i\omega)e^{-2i\omega
x_{1}}}{(V_{1}-i\omega)(V_{2}-i\omega)-V_{1}V_{2}e^{2i\omega(x_{1}-x_{2})}},
\label{AR2}\ee \be {\tilde
R}_{2,eff}=\frac{V_{1}(V_{2}+i\omega)e^{2i\omega
x_{1}}-V_{2}(V_{1}-i\omega)e^{2i\omega x_{2}}}
{(V_{1}-i\omega)(V_{2}-i\omega)-V_{1}V_{2}e^{2i\omega(x_{1}-x_{2})}},
\label{ATR2}\ee while the effective TC is \be
T_{2,eff}=A_{4}=\frac{-\omega^{2}}{(V_{1}-i\omega)(V_{2}-i\omega)-V_{1}V_{2}e^{2i\omega(x_{1}-x_{2})}}.
\label{AT2}\ee Thus with (\ref{RT}), we straightly have
\begin{align}
&R_{1}+\frac{R_{2}T_{1}^{2}}{1-\tilde{R}_{1}R_{2}}\nonumber\\
&=-\frac{V_{1}}{V_{1}-i\omega}e^{-2i\omega x_{1}}
+\frac{-\frac{V_{2}}{V_{2}-i\omega}\left(
\frac{-i\omega}{V_{1}-i\omega}\right) ^{2} e^{-2i\omega x_{2}}}
{1-\left( \frac{V_{1}}{V_{1}-i\omega}\right)\left( \frac{V_{2}}{V_{2}-i\omega}\right) e^{2i\omega(x_{1}-x_{2})}}\nonumber\\
&=\left(-\frac{1}{V_{1}-i\omega} \right) \left[V_{1} e^{-2i\omega
x_{1}} +\frac{V_{2}(-i\omega)^{2}e^{-2i\omega x_{2}}}
{(V_{1}-i\omega)(V_{2}-i\omega)-V_{1}V_{2}e^{2i\omega(x_{1}-x_{2})}}\right] \nonumber\\
&={ R}_{2,eff},
\end{align}

\begin{figure}[htbp]
    \includegraphics[scale=2,width=0.48\textwidth]{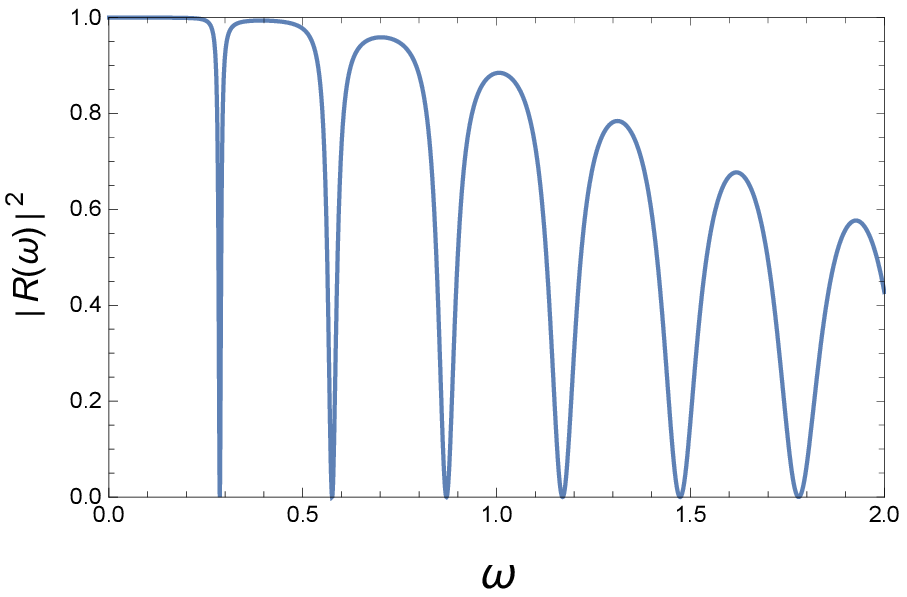}
    \includegraphics[scale=2,width=0.48\textwidth]{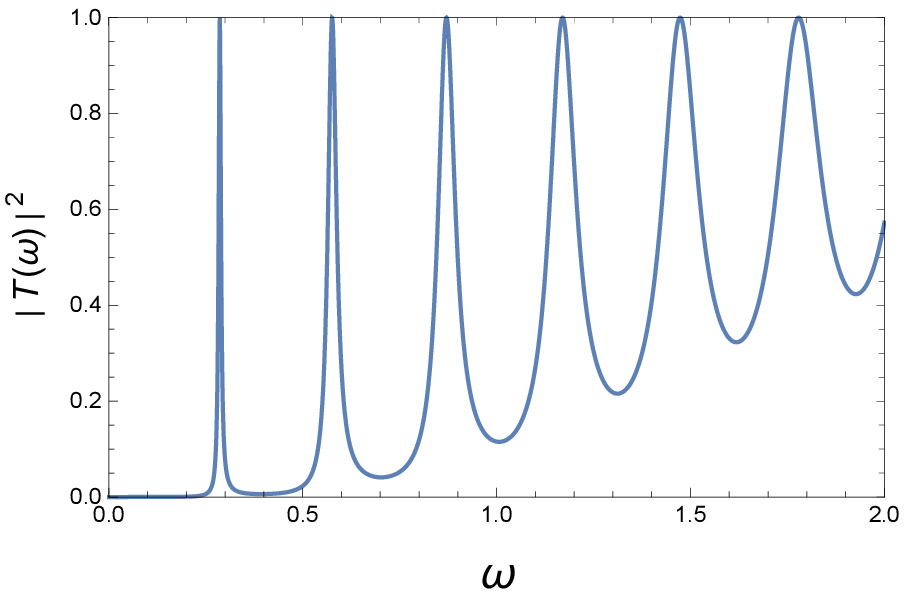}
    \caption{$ |{R}_{2,eff}|^{2} $ and $ |{T}_{2,eff}|^{2} $ with parameters $ V_{1}=V_{2}=0.5 $, $ x_{1}=-10 $ and  $ x_{2}=-20 $ .}
    \label{}
\end{figure}

Similarly, the effective RC for $N=3$ is
\begin{align}
{ R}_{3,eff}=\frac{P}{Q}
\end{align}
where \begin{align}
P&=-V_{1}(V_{2}-i\omega)(V_{3}-i\omega)e^{-2i\omega x_{1}}
\nonumber\\
&+V_{2}(V_{1}-i\omega)(V_{3}-i\omega)e^{-2i\omega x_{2}}
-V_{3}(V_{1}+i\omega)(V_{2}+i\omega)e^{-2i\omega x_{3}}
+V_{1}V_{2}V_{3}e^{2i\omega(x_{2}-x_{1}-x_{3})},\\
Q&=(V_{1}-i\omega)(V_{2}-i\omega)(V_{3}-i\omega)\nonumber\\
&-V_{1}V_{2}(V_{3}-i\omega)e^{2i\omega(x_{1}-x_{2})}
+V_{1}V_{3}(V_{2}+i\omega)e^{2i\omega(x_{1}-x_{3})}
-V_{2}V_{3}(V_{1}-i\omega)e^{2i\omega(x_{2}-x_{3})}.
\end{align}
Considering (\ref{AR2}), (\ref{ATR2}) and (\ref{AT2}), we have
\begin{align}
&{R}_{2,eff}+\frac{R_{3}{T}_{2,eff}^{2}}
{1-\tilde{R}_{2,eff}R_{3}}\nonumber\\
&=\frac{V_{2}(V_{1}+i\omega)e^{-2i\omega x_{2}}
-V_{1}(V_{2}-i\omega)e^{-2i\omega x_{1}}}
{(V_{1}-i\omega)(V_{2}-i\omega)-V_{1}V_{2}e^{2i\omega(x_{1}-x_{2})}}
+\frac{\left( -\frac{V_{3}}{V_{3}-i\omega}e^{-2i\omega
x_{3}}\right)
\left[\frac{-\omega^{2}}{(V_{1}-i\omega)(V_{2}-i\omega)-V_{1}V_{2}e^{2i\omega(x_{1}-x_{2})}}
\right]^{2} } {1-\frac{V_{1}(V_{2}+i\omega)e^{2i\omega
x_{1}}-V_{2}(V_{1}-i\omega)e^{2i\omega x_{2}}}
{(V_{1}-i\omega)(V_{2}-i\omega)-V_{1}V_{2}e^{2i\omega(x_{1}-x_{2})}}
\left( -\frac{V_{3}}{V_{3}-i\omega}e^{-2i\omega x_{1}}\right)}\nonumber\\
&=\frac{\left[V_{2}(V_{1}+i\omega)e^{-2i\omega x_{2}}
-V_{1}(V_{2}-i\omega)e^{-2i\omega x_{1}}
\right]Q-\omega^{4}V_{3}e^{-2i\omega x_{3}}} {{\left[
(V_{1}-i\omega)(V_{2}-i\omega)-V_{1}V_{2}e^{2i\omega(x_{1}-x_{2})}\right]
Q}} \nonumber\\ &={ R}_{3,eff}\label{efR3}.
\end{align}

\begin{figure}[htbp]
    \includegraphics[scale=2,width=0.48\textwidth]{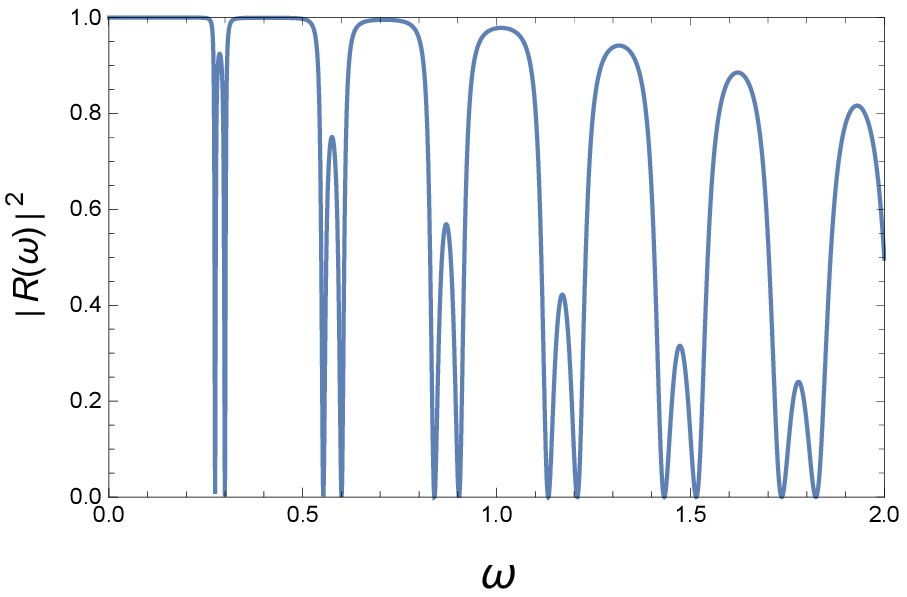}
    \includegraphics[scale=2,width=0.48\textwidth]{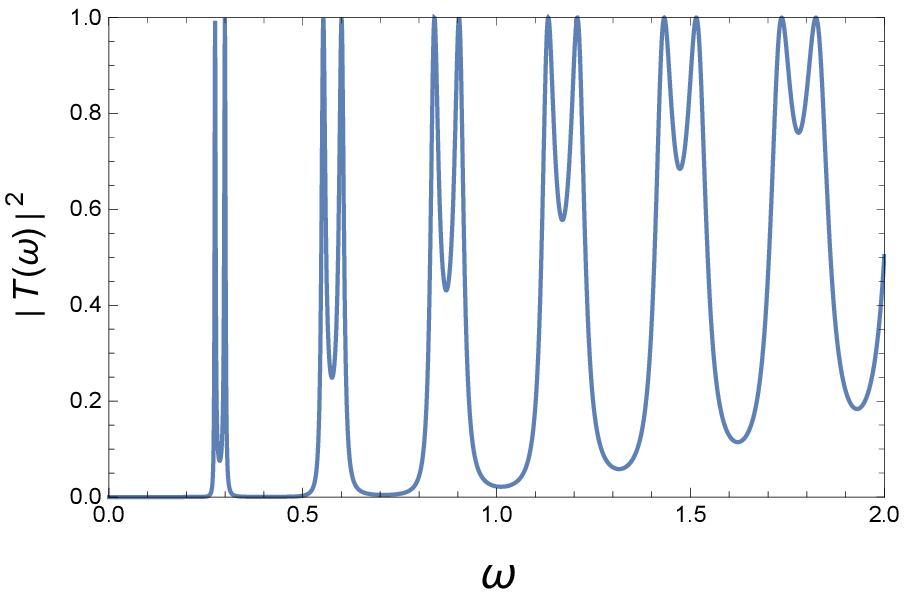}
    \caption{$ |{R}_{3,eff}|^{2} $ and $ |{T}_{3,eff}|^{2} $ with parameters $ V_{1}=V_{2}=V_{3}=0.5 $, $x_{1}=-10$, $x_{2}=-20 $ and $x_3=-30$.}
    \label{}
\end{figure}

 \end{document}